\documentclass[3p,review]{elsarticle}
\usepackage{hyperref,amsmath}
\usepackage[normalem]{ulem}
\usepackage{xcolor}
\journal{Journal of Magnetism and Magnetic Materials}

\begin{document}

\begin{frontmatter}

\title{Dynamics and morphology of chiral magnetic bubbles in perpendicularly 
magnetized ultra-thin films}

\author[ISI,Polito]{Bhaskarjyoti Sarma}


\author[INRIM]{Felipe Garcia-Sanchez}

\author[ISI,Polito]{S. Ali Nasseri}

\author[INRIM]{Arianna Casiraghi}

\author[ISI,INRIM]{Gianfranco Durin}

\address[ISI]{ISI Foundation, Via Chisola 5, 10126, Torino, Italy}
\address[Polito]{Dep. of Applied Science and Technology, Politecnico di Torino, 
Corso Duca degli Abruzzi, 24, 10129 Torino, Italy}
\address[INRIM]{Istituto Nazionale di Ricerca Metrologica, Str. delle Cacce, 
91, 10135 Torino, Italy}


\begin{abstract}
We study bubble domain wall dynamics using micromagnetic simulations in 
perpendicularly magnetized ultra-thin films with disorder and 
Dzyaloshinskii-Moriya interaction. Disorder is incorporated into the material 
as grains with randomly distributed sizes and varying exchange constant at the 
edges. As expected, magnetic bubbles expand asymmetrically along the axis of 
the in-plane field under the simultaneous application of out-of-plane and 
in-plane fields. Remarkably, the shape of the bubble has a ripple-like part 
which causes a kink-like (steep decrease) feature in the velocity versus 
in-plane field curve.  We show that these ripples originate due to the 
nucleation and interaction of vertical Bloch lines. Furthermore, we show that 
the Dzyaloshinskii-Moriya interaction field is not constant but rather depends 
on the in-plane field. We also extend the collective coordinate model for domain wall motion to a magnetic bubble and compare it with the results of micromagnetic simulations.
\end{abstract}


\end{frontmatter}

%\linenumbers

\section{Introduction}

The study of domain wall (DW) dynamics in ultra-thin films and nanowires has 
attracted significant attention in the spintronics research community due to 
its potential for applications in future memory \cite{PAR-08,HAY-08,WOL-06}, 
logic \cite{ALL-05} and sensing \cite{WEI-13,BOR-17} devices. All these 
applications require moving multiple DWs precisely with applied spin-polarised 
currents or magnetic fields. Initially, DW dynamics in Permalloy with in-plane 
magnetic anisotropy were extensively studied 
\cite{HAY-07,LEP-09,KLA-05a,YAM-04,MEI-07}. Afterwards, 
perpendicularly magnetized ultra-thin films attracted particular interest due to 
narrower domain walls compared to their in-plane magnetized counterparts. It was 
found that current-driven DW motion provides higher efficiency due to the 
enhanced values of spin-torque efficiency \cite{ALV-10,BOU-08}, with the 
DWs moving in the same direction as that of the electrons flow. On the contrary, 
in heterostructures composed of a magnetic ultra-thin film adjacent to a heavy 
metal layer, it was found that DWs move in the direction opposite to the flow of 
electrons. This behaviour was attributed to the spin Hall effect \cite{EMO-13}, 
which acts on the walls having a N\'{e}el configuration. In simultaneous 
developments, it was suggested that in such ultra-thin films the 
Dzyaloshinskii-Moriya Interaction (DMI) \cite{DZY-58,MOR-60} can 
result in a N\'{e}el wall type rather than a Bloch one 
\cite{THI-12,KIM-13,RYU-13,LEE-14}. Interfacial DMI arises 
due to the high spin-orbit coupling in the heavy metal layer and the broken 
inversion symmetry along the thin-film layers. It results in an anti-symmetric 
exchange interaction, favoring an orthogonal orientation between two neighboring 
spins in contrast with the parallel alignment of the Heisenberg exchange 
interaction. Remarkably, the DMI imposes the magnetization to rotate from one 
domain to the next with preferred handedness or chirality, resulting in 
right-handed and left-handed chiral N\'{e}el DWs. Such a property is essential 
for the spin-orbit torque to drive the DWs in the same direction, a feature that 
is fundamental for the realization of the next generation of devices, such as 
racetrack memories. Furthermore, two nearby N\'{e}el DWs with opposite 
chirality are extremely stable topologically, thus making them particularly 
suitable for applications.

In order to precisely control future spintronics devices, it is imperative to 
understand, control and measure the DMI. Several efforts have been 
made to estimate its value using different methods: asymmetric magnetic bubble 
expansion \cite{JE-13,HRA-14}, magnetic stripe domains annihilation 
\cite{JAI-17}, and Brillouin light scattering \cite{SOU-16}. Especially for 
low values, the estimation can vary dramatically from method to method, also 
for nominally identical material systems. This is probably due to the strong 
sensitivity to interface quality, growth conditions \cite{KHA-16,LAV-15}, 
thickness of heavy metal layer \cite{TAC-17} etc. 
The simplest and thus most widely used method relies on magneto-optical 
measurements of the asymmetric expansion of a magnetic bubble under the 
simultaneous presence of in-plane and out-of-plane magnetic fields. In these 
experiments, the DMI value is inferred from the measure of the DMI field, i.e. 
the in-plane field at which the velocity of the bubble DW reaches a minimum. 
Implicitly, the DW is assumed to keep its width fixed during the expansion, 
which results in a constant DMI field.

In this paper, we aim to test these implicit assumptions with micromagnetic 
simulations of the DW dynamics in a disordered medium, where the disorder is 
realistically incorporated into the material as a collection of grains with 
randomly distributed sizes and varying exchange constant at the edges. 
Significantly, we reveal that the width of the DW, and consequently the value 
of the DMI field, are not constant, as both depend on the strength of the 
applied in-plane field. Furthermore, we find that during expansion the 
nucleation of vertical Bloch lines takes place, dramatically influencing the bubble
morphology, as they move, interact and annihilate. As a consequence, the bubble 
does not grow anymore and flattens, causing a kink-like (sharp decrease) 
feature in the velocity vs. in-plane field curve.

\section{Methods}

\begin{figure}[tbt]
\centering	
\includegraphics[width=15cm]{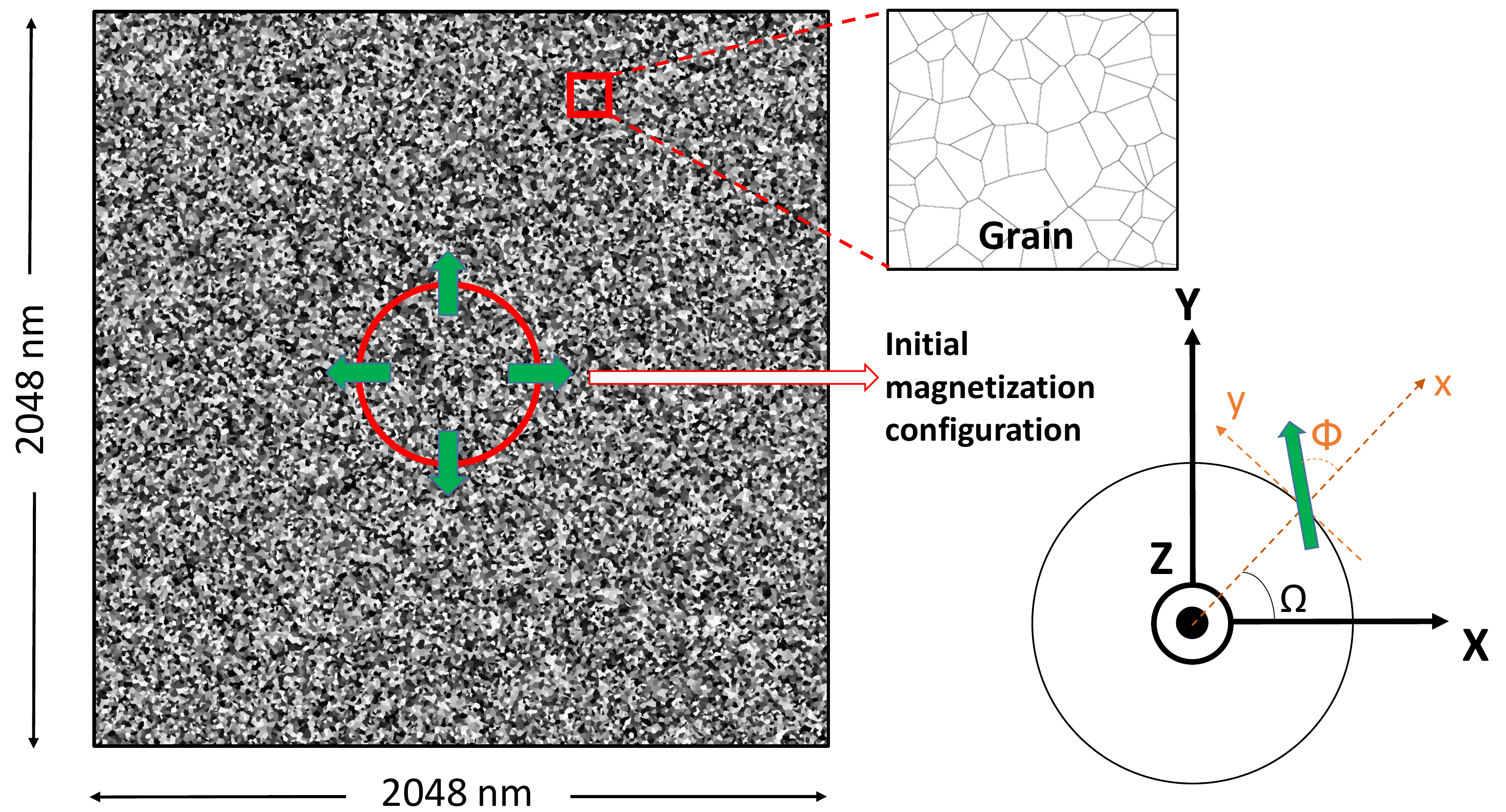}
\caption{Schematic of the simulation system depicting its dimension, initial 
magnetization configuration and grains. The thickness of the system is 0.6 nm. 
The exchange constant is reduced at the border of the grains by 43$\%$. }
\label{fig:system}
\end{figure}

Micromagnetic 2D simulations are performed using the software package Mumax$^3$ 
\cite{VAN-14,LEL-18} in a system of 1024 x 1024 x 1 rectangular cells of size 2 
nm x 2 nm x 0.6 nm, as schematically shown in 
Fig.~\ref{fig:system}. This system mimics a $Pt/Co_{90}Fe_{10}/Pt$ \cite{KIM-09} 
ultra-thin square film of 2 x 2 $\mu$m, with material constants of saturation 
magnetization $M_s=~1353~kA/m$, perpendicular magnetic anisotropy $K_u = 1.5~ 
MJ/m^3$, exchange constant $A_{ex} = 13~pJ/m$, and a Gilbert damping parameter 
$\alpha = 0.2$. The DMI values used (0.3, 0.5, 0.75, and 1 $mJ/m^2$) correspond 
to different thicknesses of the Pt layers. The material disorder is 
realistically simulated \cite{LEL-14,LEL-14b} by a random distribution of grains of 
average size of 10 nm, and an exchange constant being varied at the border of 
the grains by 43$\%$. 
A bubble domain of radius 256 nm is initialized in the center of the system and 
then allowed to expand under the simultaneous application of out-of-plane and 
in-plane fields. Its initial magnetization configuration is the minimum energy 
configuration chosen out of different possible configurations. The bubble 
domain is allowed to expand till it almost reaches the boundary of the system. 
To realistically simulate a large system, we correct for the dipolar energy to 
that of an infinite system by adding the field generated from outside the 
simulated square. 

\section{Results}

\subsection{Dynamics of the bubble domain}
\begin{figure}[tbt]
\centering
\includegraphics[width=15cm]{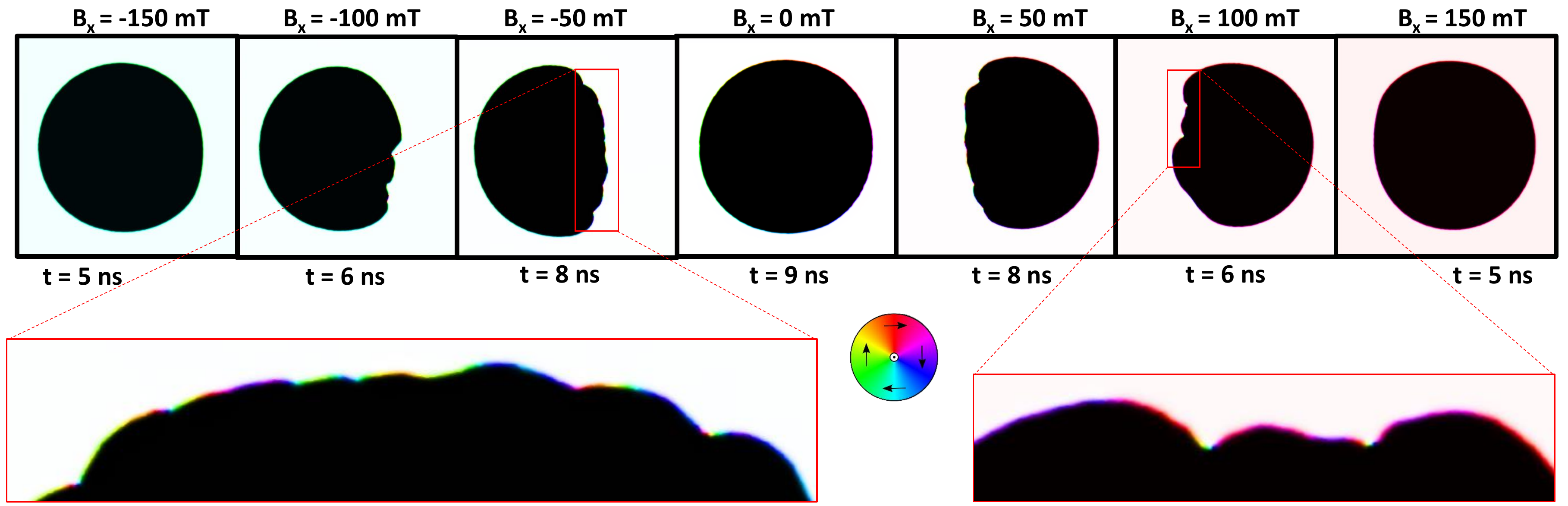}
\caption{Snapshots of a bubble domain during its expansion under the application of an out-of-plane field of $-17~mT$ and an in-plane field ($B_x$) as indicated, for a DMI value of $0.5~mJ/m^2$. Black means magnetization into the plane and white means magnetization out of 
the plane. The wheel represents the color code for in-plane magnetization direction, red meaning magnetization along positive x-direction and blue meaning magnetization along negative y direction.}
\label{fig:bubbles}
\end{figure}

In Fig.~\ref{fig:bubbles} we present a snapshot of the bubble domain dynamics 
for a DMI constant of $0.5~mJ/m^2$ during its evolution under the application of 
an out-of-plane field of $-17~mT$ for different in-plane fields. Without an 
in-plane field the bubble domain expands symmetrically, while for a non-zero 
in-plane field, it becomes strongly asymmetrical, as expected. Remarkably, for 
in-plane fields between $30~ mT$ and $100~ mT$, the front opposite to the 
direction of the in-plane field becomes “ripple-like”, and the magnetization 
inside the DW undergoes a complex rotation as we show in the zoomed-in areas of 
Fig.~\ref{fig:bubbles}. On the other hand, for very low and for very high 
in-plane fields, these ripples are absent and the domain wall appears rather 
smooth. 

\begin{figure}[tbt]
\centering
\includegraphics[width=15cm]{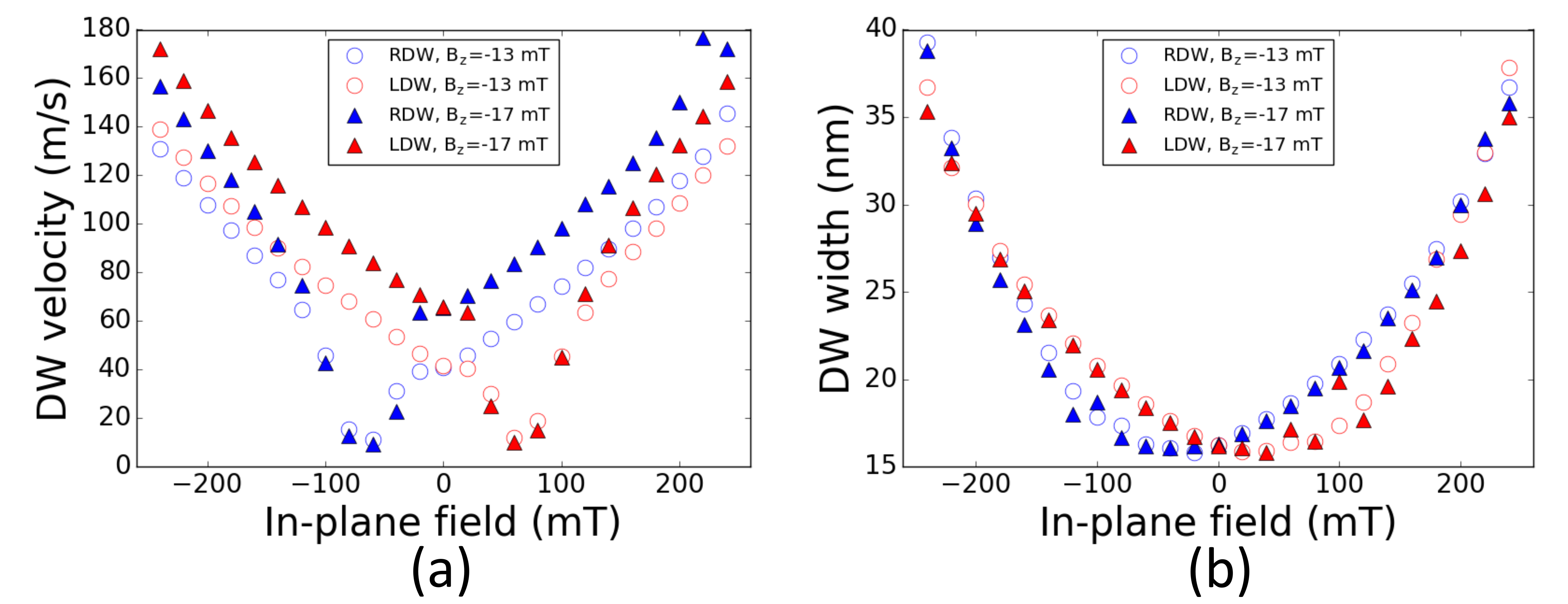}
\caption{Velocity (a) and width (b) of the right domain wall (RDW) in blue and 
left domain wall (LDW) in red as a function of in-plane field for $D=0.5~mJ/m^2$. Circles (empty) and triangles (solid) represent velocities at out-of-plane 
fields of $-13$ and $-17~mT$, respectively.}
\label{fig:vel_width}
\end{figure}

In order to understand how the ripples affect the bubble expansion, we show in 
Fig.~\ref{fig:vel_width}(a) the velocity of the right domain wall (RDW) and 
left domain wall (LDW) as a function of in-plane fields at two different 
out-of-plane fields ($-13~mT$ and $-17~ mT$). As the in-plane field decreases 
from higher positive values towards zero, the velocity of the RDW decreases. It 
keeps decreasing till a negative in-plane field value, then steeply decreases, 
reaches its minimum and then increases again. For comparison, we show the width 
of the domain wall as a function of in-plane fields in 
Fig.~\ref{fig:vel_width}(b) at the same out-of-plane fields. For both cases, the 
minimum is not at zero in-plane field, but it is shifted at values roughly 
corresponding to the onset of the drop in the DW velocity. On the other hand, 
it does not significantly depend on the out-of-plane fields.

\begin{figure}[tbt]
\centering
\includegraphics[width=15cm]{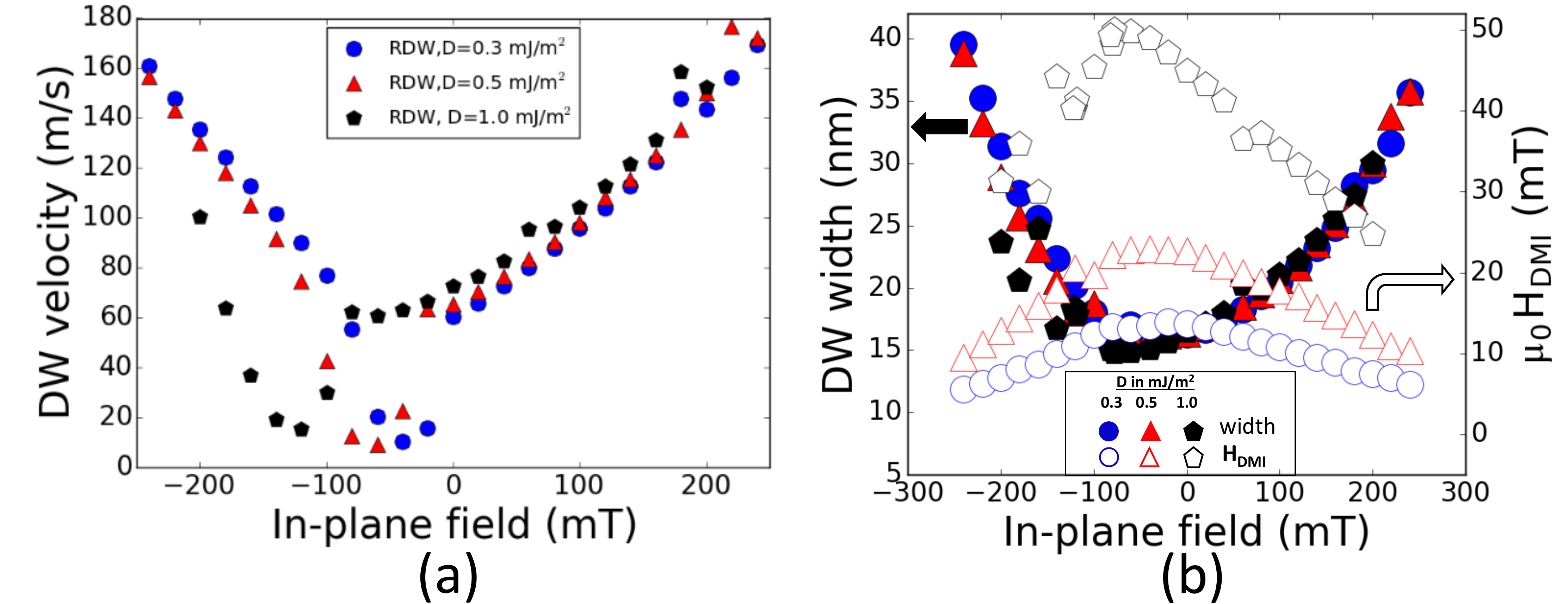}
\caption{(a) Velocity of the right domain wall (RDW) as a function of in-plane 
field at constant out-of-plane field of $-17~mT$, (b) width and DMI field of 
the RDW as a function of in-plane field for 3 different DMI constants keeping 
the OOP field constant at $-17~mT$.}
\label{fig:vel_width_D}
\end{figure}

Fig.~\ref{fig:vel_width_D}(a) displays the velocity of the RDW as a function of 
in-plane field for different DMI constants, and an out-of-plane field of 
$-17~mT$. The nature of the velocity vs. in-plane field curve is similar to 
Fig.~\ref{fig:vel_width}(a), moving the minimum at larger (negative) values for 
increasing DMI value. Correspondingly, we show in Fig.~\ref{fig:vel_width_D}(b) 
the width of the RDW and the DMI field as a function of in-plane field. 
The DMI field $H_{DMI}$ is calculated using the expression $\mu_0 H_{DMI} = D / 
(M_s \Delta)$, where $D$ is the DMI constant and $\Delta$ is the DW width. 
The latter depends on the in-plane field and has its minimum shifted towards a 
negative in-plane field value in the same way as in Fig.~\ref{fig:vel_width}(b), 
while it does not show any marked dependence on the DMI constant. Having a 
dependence on the DW width, we can conclude that the DMI field also varies 
with the in-plane field. 

A few characteristic points can be identified in the velocity curves of 
Figs.~\ref{fig:vel_width}-~\ref{fig:vel_width_D}, 
as reported in Table~\ref{tab:fields}. By 'Onset' we mean the in-plane field 
at which the velocity of the wall decreases steeply and the 'Minimum' is the 
in-plane field at which the velocity of the wall is minimum. In the table, 
magnitude of both the onset and minimum fields increase as a function of DMI. 

\begin{table}
  \centering
  \begin{tabular}{|c|c|c|}
  \hline
      D ($mJ/m^2$) & Onset ($mT$) & Minimum ($mT$)\\
      \hline
      0.3 & 0$\pm$2 & -40$\pm$2\\
      \hline
      0.5 & -22$\pm$2 & -62$\pm$2\\
      \hline
      0.75 & -38$\pm$2  & -100$\pm$2\\
      \hline
      1 & -80$\pm$2 & -120$\pm$2\\    
      \hline
  \end{tabular}
  \caption{Onset field and minima field for the RDW at $B_z=-17 mT$ for 
different DMI constants} 
  \label{tab:fields}
\end{table}

\subsection{Nucleation of vertical Bloch lines}
\begin{figure}[tbt]
\centering
\includegraphics[width=15cm]{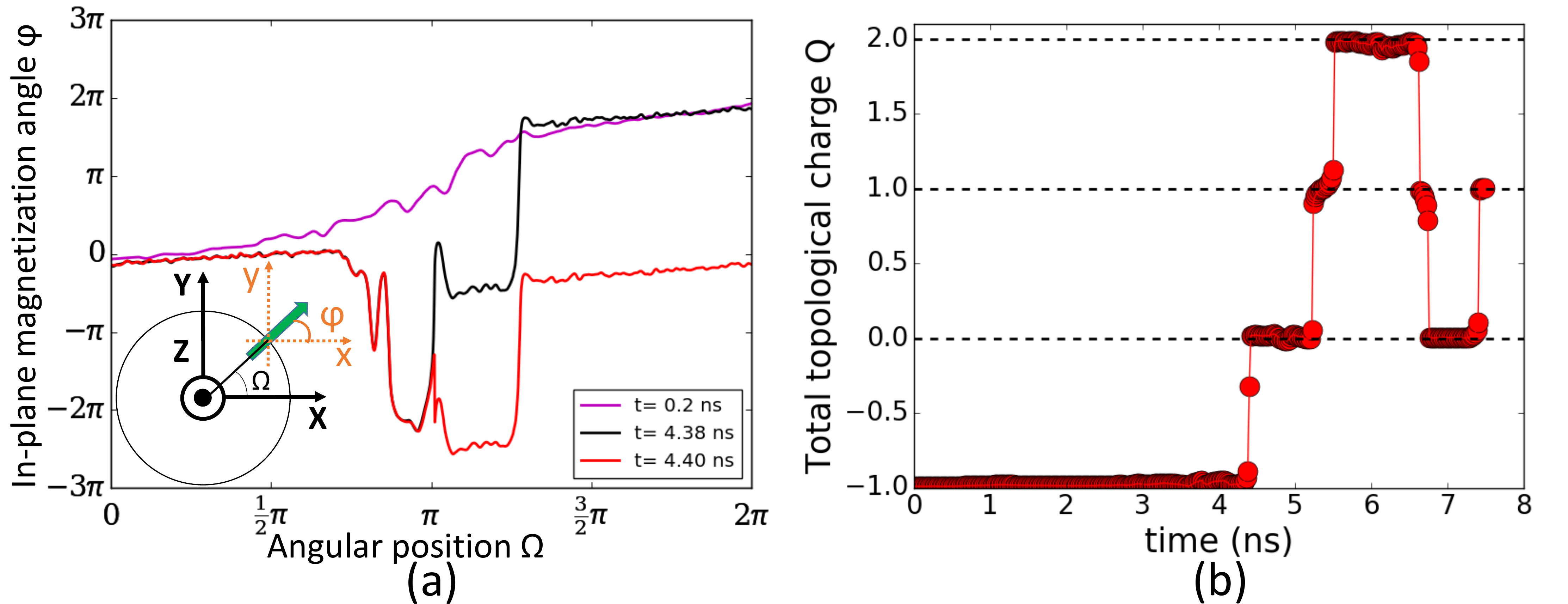}
\caption{(a) In-plane magnetization angle $\varphi$ of the domain wall along the periphery 
of the bubble as a function of angular position of the DW. Positive (negative) angle means an anti-clockwise (clockwise) rotation. 
(b) Evolution of the total topological charge $Q$ of the system showing six annihilation events of $\pm 1$, 
corresponding to annihilation of two VBLs with the same topological charge of 
$\pm 1/2$.}
\label{fig:topol}
\end{figure}

The occurrence of ripples and the flattening of the bubble domain is clearly 
related to the kink like feature in the velocity vs in-plane field curve, described above. To 
better understand its origin, we present in Fig.\ref{fig:topol}(a) the magnetization angle $\varphi$ at the centre of the DW (with φ measured with respect to a fixed coordinate system, as in the sketch) against its position around the bubble 
(expressed by the angle $\Omega$). With this representation, a perfect N\'{e}el 
wall all around the bubble periphery translates into a smooth increase from the 
coordinate $(0,0)$ to $(2\pi, 2\pi)$. On the other hand, a Bloch wall all along 
the periphery would increase from $(0, \pi/2)$ to $(3/2\pi, 
2\pi)$  and then decrease to $(2\pi, \pi/2)$. 

We present the case for an out-of-plane field of $-17~mT$, an in-plane field of $100~mT$ and a DMI constant of 
$0.5~ mJ/m^2$ as a typical behavior. At 0.2 ns after the application of the fields, the 
in-plane magnetization rotates along the periphery from 0 to $2\pi$ with 
some local fluctuations, showing a clear N\'{e}el wall configuration. At 4.38 
ns, the magnetization still starts at 0 and ends at $2\pi$,
but showing big fluctuations. In particular, the in-plane magnetization suddenly 
makes a $2\pi$ rotation in clockwise direction (negative), comes back to zero with 
an anticlockwise rotation of $2\pi$ and makes another $2\pi$ rotation 
anticlockwise to end at $2\pi$. A rotation of $2\pi$ angle of the in-plane 
magnetization along the periphery corresponds to the onset of a pair of vertical 
Bloch lines (VBL) \cite{YOS-15}. As a matter of fact, three pairs of VBLs occur at
t = 4.38 ns and two pairs at t=4.40 ns, meaning that two VBLs annihilate 
during the 20 ps time interval. These complex rotations of $2\pi$ angle are 
also visible in Fig.~\ref{fig:bubbles} as color fluctuations in the bubble 
front.

The VBLs are nucleated in disordered systems due to the incoherent 
precession of the magnetic moments within the DWs due to the spatial inhomogeneities of the effective field. 
Clearly, they are nucleated in pairs in order to conserve the total 
topological charge $Q_{total}$ of the system. In other words, they must have opposite topological charge $\pm Q_{VBL}$.
After nucleation, these VBLs start propagating and interact with each 
other. Different VBLs can have different widths 
due to the inhomogeneous component of the $B_x$ acting on them, so that 
they also have different velocities \cite{YOS-15}. As a consequence, fast VBLs 
come close to the slower ones and can annihilate each other if energetically 
favorable. These annihilation events can be studied in terms of time evolution 
of the total 
topological charge $Q_{total}(t)$, expressed as \cite{BRA-12a}:

\begin{equation}
 Q_{total}(t) = \frac{1}{4\pi} \iint \textbf{m}(t) \left( \frac{\partial \textbf{m}(\textbf{r},t)}{\partial x} \times  \frac{\partial \textbf{m}(\textbf{r},t)}{\partial y} \right) dx dy
\end{equation}

where $\textbf{m}(t)$ is the normalized magnetization at time $t$ and position $\textbf{r}$. 
Fig.~\ref{fig:topol}(b) shows six annihilation events where the topological 
charge jumps by $\pm 1$, for which a pair of VBLs annihilate both having topological charge 
+1/2 or -1/2 respectively. Clearly, these jumps are not perfectly an integer number, due to the presence of disorder.

\subsection{Comparison with models of DW dynamics}

\begin{figure}[tbt]
\centering
\includegraphics[width=15cm]{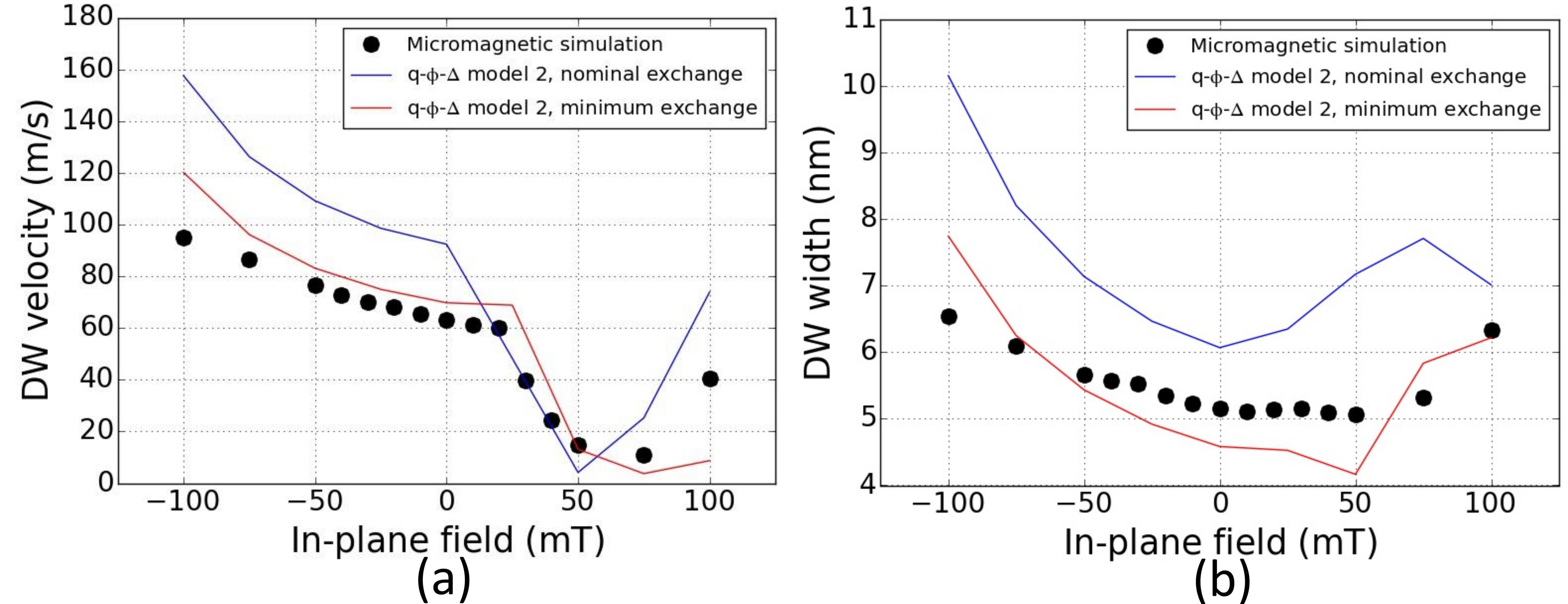}
\caption{Comparison between the micromagnetic simulations and collective coordinate model for the LDW at a drive field of $B_z = -17~mT$. Nominal exchange is the actual value of the exchange constant and minimum exchange is the exchange constant after reducing it by 43\% of the actual exchange constant.}
\label{fig:model}
\end{figure}

While the analysis above explains the drop and the mimimun of the DW 
velocity of Fig.~\ref{fig:vel_width}, it seems to break the interpretation of 
the DW dynamics in simple terms, as for 1D class model \cite{THI-12}, for which 
a linear relation between velocity and DW width exists. In particular, while 
the velocity shows a significant dependence on the out-of-plane field, the width 
is totally unaffected. On the other hand, a failure of simple 1D models is 
totally expected in case of VBLs nucleation and annihilation. In 1D models, in 
fact, the magnetic moments inside the DW are assumed to vary only along one 
dimension and this is not the case when VBLs are present. 
To account for these discrepancies, we extended collective 
coordinate models (CCMs) that go beyond the simple 1D models \cite{NAS-17b},
to the bubble dynamics considered here. For simplicity, we assume that the 
points on the bubble are free and not interacting with each other. We then
introduce a local in-plane field $H_{eff}$, and a local dipolar field (with 
demagnetizing factor $N$) at points on the bubble periphery as follows:

\begin{flalign} 
\label{eq:bubble}
H_{x, eff} &= H_X cos(\Omega) + H_Y sin (\Omega) \nonumber\\
H_{y, eff} &= -H_X sin (\Omega) + H_Y cos (\Omega) \\
N_{x, eff} &= N_X cos (\Omega) + N_Y sin (\Omega) \nonumber\\
N_{y, eff} &= -N_X sin (\Omega) + N_Y cos (\Omega) \nonumber
\end{flalign}
where $x$ and $y$ are the local axes over which the equations are written, $X$ and $Y$ are the global axes, as in Fig.~\ref{fig:system}. With these assumptions, 
the equations derived for the CCMs will be exactly those found in 
Ref.~\cite{NAS-17b}, with no DW tilting and the local fields above replacing 
the global values.
We compared this model to the results of micromagnetic simulations to assess 
whether these models are accurate especially considering the disorder included 
in the model. As depicted in Fig.~\ref{fig:model}. we find that by reducing the exchange constant by 43\% of the nominal value (equivalent to the amount of exchange constant variation at the grain boundaries) we are able to almost reproduce the micromagnetic results. As such, the toy model seems to be 
valid at least for cases of low drive field.

\section{Discussion}

The fact that the magnetic bubble expands symmetrically without an applied 
in-plane field and asymmetrically when a non-zero in-plane field is applied, as 
shown in Fig.~\ref{fig:bubbles}, is well-known for most perpendicular 
ultra-thin films with DMI. In our simulations we use positive values of DMI, so that the DMI field acts on the bubble domain in the radially outward direction. A 
positive in-plane field is thus parallel to the DMI field on the right side of 
the bubble and antiparallel on the left side. When these fields are parallel 
(antiparallel) they stabilize (destabilize) the DW. This simple picture helps us 
to understand why it is possible for the ripples to form. Incidentally, this is 
valid only when the two fields are comparable in amplitude, as at 
higher in-plane fields the DWs are very stable again.

As shown, the formation of ripples is reflected in the velocity curves in 
Figs.~\ref{fig:vel_width}(a),  and \ref{fig:vel_width_D}(a). Some aspects of 
these curves can be understood with the help of the equation for free energy of 
the domain wall, given by \cite{JUE-16}:
\begin{equation}
 \sigma_{DW}(H_x, \Phi) = \sigma_0 + 2 K_D \Delta cos^2(\Phi) - \pi\Delta M_s 
(H_x+H_{DMI}) cos(\Phi)
\end{equation}

where $\sigma_0$ is the Bloch wall energy, and $K_D$ the domain wall 
anisotropy energy. This equation tells us that the energy of the DW is 
anisotropic and depends on the magnetization angle of the DW. The energy of the 
DW is then maximum when $H_x$ and $H_{DMI}$ are antiparallel and minimum when 
parallel. The DWs with the maximum energy have the minimum velocity. This 
expression explains why there is a horizontal shift in the velocity curve, but cannot 
explain the asymmetry, relative to the curves minima, observed in our simulations as well as in other 
experiments \cite{LAV-15}. It has been speculated that a chiral damping arising 
out of spin-orbit interaction could be responsible for this \cite{JUE-16}. 
D.-Yun Kim \textit{et al.} proposed that the asymmetry in the velocity curve is due to the 
dependence of energy on DW width \cite{KIM-16}. In order to understand the 
flattening of the bubble, we need to explain it taking into account the formation 
of VBLs that we observe. Although the flattening of the bubble \cite{LAU-16} or 
kink-like feature \cite{LAV-15} of velocity vs in-plane field curve has been 
observed experimentally, most of these works focus on extracting the value of 
the DMI, while the shape of the bubble is seldom studied. D. Lau \textit{et al.} 
used energetic calculations of the equilibrium shape of the bubble domain wall 
by Wulff construction in order to explain the shapes \cite{LAU-16}. In their 
studies, they observed an asymmetric expansion of the bubble with a flattening 
DW at lower in-plane fields and a non-elliptical tear-drop shape at higher 
in-plane fields. While the tear-drop shape could be explained using the Wulff 
construction, it was not straightforward to explain the flattening shape. It was 
speculated in their work that the flattened shape could be due to the 
nucleation of vertical Bloch lines. Our detailed study of the rippled points 
confirm that the flattening is due to the nucleation and interaction of Vertical 
Bloch lines. The nucleation of VBLs has been observed for Co/Ni 
wide strips experimentally \cite{YOS-15} and is predicted to occur when 
$H_{DMI}$ is antiparallel to $H_x$. When the DMI is stronger, a higher in-plane 
field is needed for the formation of VBLs to take place. This explains why the 
onset and minimum points in table~\ref{tab:fields} are higher for higher DMI 
values. VBLs are high energy regions in the DW and therefore sections of DWs 
having VBLs have smaller velocities compared to the rest of the DW. 
Morphologically, they appear as pinned DW points, resulting in the occurrence 
of the ripples. VBLs are then responsible for reducing the overall velocity of 
one side of the bubble DW. Furthermore, we expect the velocity of the DW to be 
inversely proportional to the density of VBLs. Different velocities at 
different in-plane fields, as shown in Fig.~\ref{fig:vel_width}(a), are related 
to the difference in the density of VBLs. The velocities of the right and left DWs have also been predicted by Pellegren et al. \cite{PEL-17} using a dispersive elastic stiffness model. In this model, the velocities are calculated from a modified DW elastic energy scale using the creep law. For small length scales, as in the case of our system, this model predicts the onset of the drop in the DW velocity as well as the convergence of RDW and LDW velocities at higher in-plane fields. These remarkable similarities between this model, which assumes non-zero temperature using energy barrier scaling, and our simulations, which are instead simulations of the dynamics performed at zero temperature, suggest that the observed properties of DW propagation under simultaneous application of in-plane and perpendicular fields originate from the intrinsic DW energy.

In addition, we can also use the CCM to understand what happens at the ripple 
points. The steady state solution ($\Phi \sim 0$) for the magnetization angle 
of the DW in this case reads:
\begin{equation}
 cos(\Phi_{eq}) = \frac{1}{2} \left[ \frac{I^2_2}{I_1 I_4} 
\frac{H_z}{H_W} csc(\Phi_{eq}) + \alpha \frac{I_3}{I_4} 
\frac{H_{DMI}}{H_W} - \alpha \frac{I_6}{I_4} 
\frac{H_{x,eff}-H_{y, eff} cot(\Phi_{eq})}{H_W} \right]
\end{equation}

where $\Phi_{eq}$ is the steady state magnetization, $H_W$ is the conventional 
Walker Breakdown field, and the $I_i$ are constants calculated for a specific 
Bloch profile \cite{NAS-17b}. If the equation above does not have a solution, the 
magnetization will continue to precess and can be determined using the full 
collective coordinate model. In the absence of the DMI and in-plane fields, the 
drive field determines whether Walker breakdown happens. However, in the 
presence of the in-plane field and DMI, there are two additional parameters that 
play a role in whether or not precession continues. The fact that the in-plane 
fields are locally determined means that points of precession can nucleate 
within the bubble locally, showing local Walker breakdown behavior. Physically 
speaking, such processional motion will lead to 2D effects on the bubble, 
affecting the spin texture around that point and giving rise to the ripple like 
shape of the bubble. Even though the CCM is effectively a one-dimensional model, 
such ripples can be observed in the results as well. 

\section{Conclusions}
We have studied the dynamics of chiral magnetic bubbles in perpendicular magnetic anisotropy materials using micromagnetic simulations in the presence of disorder. As expected, we observe an asymmetry in the expansion of the 
bubble in the simultaneous presence of out-of-plane and in-plane fields. There 
is a range of applied in-plane fields in which a part of the bubble shows 
ripple-like structures. These ripples cause a kink-like feature in the 
velocity of the domain wall. We confirm that the generation of ripples is due to the nucleation of vertical Bloch lines. We find that the width of the domain wall depends on the in-plane field and for the first time to our knowledge, it brings us to the remarkable conclusion that DMI field also depends on the in-plane field and is not a constant. Future studies on vertical Bloch lines can shed more light on the dynamics and shape of magnetic bubbles in ultra-thin films and its effect on the measurement of DMI. Furthermore, we extend a collective coordinate model that explains the velocity curve qualitatively well. 

\section{Acknowledgements}
This study was conducted as part of the Marie Curie ITN WALL project, which 
has received funding from the European Union's Seventh Framework Programme for 
research, technological development and demonstration under Grant Agreement No. 
608031.

\section*{References}
%\bibliographystyle{plain}
%\bibliography{dws}

\end{document}